\documentclass[pdflatex,sn-mathphys,Numbered]{sn-jnl}

\usepackage{graphicx}%
\usepackage{multirow}%
\usepackage{amsmath,amssymb,amsfonts}%
\usepackage{amsthm}%
\usepackage{mathrsfs}%
\usepackage[title]{appendix}%
\usepackage{xcolor}%
\usepackage{textcomp}%
\usepackage{manyfoot}%
\usepackage{booktabs}%
\usepackage{algorithm}%
\usepackage{algorithmicx}%
\usepackage{algpseudocode}%
\usepackage{listings}%
\usepackage{subcaption}%
\usepackage{url}%
\usepackage{tikz}%
\usetikzlibrary{arrows.meta,fit,positioning,calc,patterns,shapes.geometric}%
\usepackage{pgfplots}%
\pgfplotsset{compat=1.18}
\usepgfplotslibrary{groupplots}

\lstdefinelanguage{TypeScript}{
  keywords={async, await, const, let, var, function, return, if, else, new, import, from, export, class, this, typeof, of, in, for, while, break, continue, switch, case, default, throw, try, catch, finally, interface, type, extends, implements, public, private, readonly},
  ndkeywords={number, string, boolean, void, any, Float64Array, Int32Array, Uint32Array, Worker, MessageChannel, SharedArrayBuffer, Atomics, Promise, Array, Map, Set, console},
  keywordstyle=\color{blue!70!black}\bfseries,
  ndkeywordstyle=\color{purple!60!black},
  sensitive=true,
  comment=[l]{//},
  morecomment=[s]{/*}{*/},
  commentstyle=\color{gray!70!black}\itshape,
  stringstyle=\color{olive!60!black},
  morestring=[b]',
  morestring=[b]",
  morestring=[b]`
}
\begin{document}

\title[ParaWeb: Parallel Programming Patterns for Web Development]{ParaWeb: Parallel Programming Patterns for Web Development}

\author*{\fnm{Suejb} \sur{Memeti}}\email{suejb.memeti@bth.se}

\affil*{\orgdiv{Department of Computer Science}, \orgname{Blekinge Institute of Technology}, \orgaddress{\city{Karlskrona}, \country{Sweden}}}

\abstract{Modern web applications increasingly require computationally intensive processing, yet JavaScript, the dominant language of the web, has traditionally been limited to a single-threaded execution model. Node.js Worker Threads and browser Web Workers provide low-level mechanisms for parallel execution, but developers lack high-level abstractions that capture recurring parallel structures as reusable patterns. In this paper, we present ParaWeb, a TypeScript library that implements ten parallel programming patterns for server-side Node.js, client-side browser environments, and WebGPU compute shaders. ParaWeb provides three implementation variants for each pattern: a message-passing (MP) variant based on structured cloning via \texttt{postMessage}, a shared-buffer (Shared) variant that uses \texttt{SharedArrayBuffer} with typed array views, and a GPU variant that uses WebGPU compute shaders for hardware-accelerated execution. We describe the architecture, design decisions, and pattern-specific implementation strategies, and we evaluate the performance of all thirty implementations across three data sizes. Experimental evaluation results show that the CPU-based variants achieve speedups of up to 11.6x with 16 threads for compute-bound patterns, while the GPU variants reach speedups of up to 260x for compute-bound patterns with high arithmetic intensity such as Farm, Scatter, Reduce, and Map. A case study on five image-convolution filters further shows that GPU acceleration reaches up to 414x speedup over single-threaded CPU on non-separable kernels, with consistent scaling across 1024$^2$, 2048$^2$, and 4K images.}


\maketitle

\section{Introduction}\label{sec:intro}

Modern web applications increasingly handle workloads that were traditionally reserved for desktop or server-side high-performance computing environments. Tasks such as image processing, real-time data analytics, scientific visualization, and machine learning inference~\cite{tfjs2019} now routinely execute within web browsers and Node.js server processes. However, JavaScript, the primary programming language of web development, was originally designed around a single-threaded event loop model, which limits its ability to exploit the multi-core CPUs and GPUs that are common in today's hardware. While this model simplifies concurrency reasoning and avoids many classes of synchronization errors, it leaves significant computational resources unused when applications encounter CPU-bound workloads.

The introduction of Worker Threads in Node.js\footnote{\url{https://nodejs.org/api/worker_threads.html}} and Web Workers in browser environments\footnote{\url{https://www.w3.org/TR/workers/}} has made true parallelism possible within the JavaScript ecosystem. These mechanisms allow developers to spawn separate execution contexts that run concurrently on different CPU cores. Supporting both server-side and client-side parallelism is important for several reasons: it avoids overloading servers by offloading computation to client devices, and it allows privacy-sensitive workloads, such as medical image processing or personal data transformations, to remain on the client when appropriate. However, these low-level primitives require developers to manually manage thread creation, data partitioning, message passing, synchronization, and result aggregation, which is a non-trivial and error-prone programming task.

The concept of \emph{algorithmic skeletons}, originally introduced by Cole~\cite{cole1989algorithmic}, provides a structured approach to expressing parallelism through high-level patterns such as Map, Reduce, Pipeline, and Farm. Several mature frameworks in the C++ ecosystem, such as SkePU~\cite{skepu3}, FastFlow~\cite{fastflow2010}, and GrPPI~\cite{grppi}, have shown that pattern-based parallel programming improves programmer productivity while providing portable performance across different hardware platforms. However, to the best of our knowledge, no pattern-based parallel programming library exists for the JavaScript ecosystem that covers Node.js Worker Threads, browser Web Workers, and WebGPU compute shaders together with multiple implementation strategies. The closest peer is River Trail~\cite{herhut2013rivertrail}, which was an engine-level effort and has been discontinued.

In this paper, we present ParaWeb\footnote{Source code: \url{https://github.com/sm13294/paraweb-js}}, a TypeScript library that implements ten parallel programming patterns for web development. The patterns are organized into three categories: (1) data-parallel patterns, including Map, Filter, Reduce, Scan, and Scatter; (2) task-parallel patterns, including Farm, Pipeline, and Divide-and-Conquer; and (3) specialized patterns, including MapReduce and Stencil. For each pattern, we provide three implementation variants: a message-passing (MP) variant that distributes data via structured cloning through \texttt{postMessage}, a shared-buffer (Shared) variant that uses \texttt{SharedArrayBuffer} with \texttt{Float64Array} views to allow zero-copy data sharing between threads, and a GPU variant that offloads computation to the graphics processor using WebGPU compute shaders written in WGSL (WebGPU Shading Language). All three variants run in modern browsers without engine changes. WebGPU is supported in Chrome~113+, Edge~113+, Firefox~121+, and Safari~18+, and the Shared variant additionally requires cross-origin isolation headers. The design of ParaWeb is inspired primarily by the GrPPI framework~\cite{grppi}, adapted for the JavaScript ecosystem.

Major contributions of this paper include:
\begin{itemize}
    \item A library of ten parallel programming patterns under a single typed API, with three implementation variants per pattern (message-passing, shared-buffer, GPU) covering Node.js Worker Threads, browser Web Workers, and WebGPU compute shaders.
    \item An empirical evaluation of all thirty implementations across three input sizes and four parallel thread counts plus a sequential baseline, on Node.js and in the browser, with the same per-element workload on every variant.
    \item A browser-based case study on five image-convolution filters (Gaussian blur, box blur, sharpen, emboss, edge detection) showing GPU acceleration scaling consistently across $1024^2$, $2048^2$, and 4K images.
\end{itemize}

The rest of this paper is structured as follows. Section~\ref{sec:related} reviews related work on parallel pattern frameworks and JavaScript parallelism. Section~\ref{sec:design} presents the architecture and design decisions of ParaWeb. Section~\ref{sec:implementation} describes pattern-specific implementation details. Section~\ref{sec:setup} outlines the experimental setup. Section~\ref{sec:evaluation} reports and discusses performance results, including the workload-dependent variant-selection observations and current limitations. Section~\ref{sec:conclusion} concludes the paper and outlines future work.

\section{Related Work}\label{sec:related}

ParaWeb builds on prior research in two areas: algorithmic skeletons and pattern-based parallel programming libraries in established HPC ecosystems, and parallel and GPU-accelerated execution models for the web. In this section, we review each of these areas and position ParaWeb with respect to existing approaches.

\subsection{Algorithmic Skeletons and Parallel Pattern Libraries}

The idea of expressing parallelism through reusable high-level patterns has been developed for several decades. Asanovi\'{c} et al.\ introduced the ``Berkeley Dwarfs,'' a taxonomy of thirteen algorithmic motifs representing the dominant computational patterns in parallel applications~\cite{asanovic2006landscape}. Cole~\cite{cole1989algorithmic} introduced \emph{algorithmic skeletons} as a structured way of expressing parallelism through high-level patterns such as Map, Reduce, Pipeline, and Farm. Both approaches capture recurring parallel structures as reusable abstractions, reducing programmer burden while allowing portable and efficient parallel execution.

Several C++ frameworks have since shown the effectiveness of implementing parallel patterns as reusable constructs. SkePU~\cite{skepu3} provides Map, Reduce, MapReduce, Scan, and Stencil patterns with portable execution across CPUs, GPUs, and HPC clusters, using OpenMP, CUDA, and OpenCL backends, with type-safe skeleton programming added in recent versions. Muesli~\cite{muesli} is another C++ skeleton library with OpenMP, CUDA, and MPI backends, providing both data- and task-parallel skeletons. FastFlow~\cite{fastflow2010} provides streaming and task-parallel patterns such as Pipeline, Farm, and Divide-and-Conquer, using lock-free queues for high throughput. Kokkos~\cite{kokkos} addresses performance portability across many-core architectures through polymorphic memory access patterns and parallel execution abstractions such as \texttt{parallel\_for} and \texttt{parallel\_reduce}. Intel TBB~\cite{reinders2007tbb} and OpenMP~\cite{openmp} include parallel patterns implicitly through generic algorithms and compiler directives.

GrPPI (Generic Reusable Parallel Patterns Interface)~\cite{grppi} provides a single interface for patterns such as Map, Reduce, MapReduce, Stencil, Pipeline, Farm, Filter, and Divide-and-Conquer, targeting different backends including OpenMP, Intel TBB, C++ threads, and FastFlow by selecting an execution policy at compile time. GrPPI has been evaluated for stream processing~\cite{grppi_stream} and multi-core performance~\cite{grppi_performance}, showing that a small, well-chosen set of patterns is sufficient for a broad range of parallel workloads, and serves as the primary design inspiration for ParaWeb.
\subsection{Parallel and GPU Computing in Web Environments}

Parallel computing on the web has historically been limited by JavaScript's single-threaded execution model. The closest peer to ParaWeb is River Trail~\cite{herhut2013rivertrail}, which extended JavaScript with a \texttt{ParallelArray} type supporting Map, Reduce, Scan, Filter, and Scatter, executing on multi-core CPUs and GPUs via OpenCL. River Trail required engine-level modifications to SpiderMonkey and was tied to OpenCL as its GPU backend. Mozilla's complementary Parallel JS (PJs) effort~\cite{matsakis2012pjs} took a similar engine-level approach. Both projects were ultimately abandoned (River Trail's code was removed from Firefox in 2015) because shipping a parallel JavaScript runtime required every engine vendor to adopt the same engine-level changes, which did not happen, and OpenCL itself never gained traction on the web. Web Workers provide a standardized, low-level mechanism for parallel task execution in browser environments, but they lack structured abstractions; developers must manually manage message passing, data partitioning, and synchronization. Their performance scalability has been characterised in detail across browsers and CPU configurations~\cite{verdu2016scalability}, and earlier work has investigated unified distributed-and-parallel JavaScript programming models built on top of them~\cite{welc2010generic}. Several libraries, such as \texttt{parallel.js}\footnote{\url{https://github.com/parallel-js/parallel.js}} and \texttt{workerpool}\footnote{\url{https://github.com/josdejong/workerpool}}, expose limited forms of Map or SIMD-like processing, but none offer a full pattern-based abstraction covering the range of parallel patterns found in established frameworks. Node.js Cluster\footnote{\url{https://nodejs.org/api/cluster.html}} enables process-based parallelism but focuses on request-level load balancing rather than structured data or task parallelism.

For compute-intensive workloads, two complementary paths have emerged for speeding up code on the web. WebAssembly~\cite{wasm} defines a portable, low-level bytecode that browsers execute at near-native CPU speed. Recent research shows that WebAssembly performance is typically within $\sim$2x of native code on standard benchmarks, with the remaining gap attributed to register pressure and memory-management overheads~\cite{jangda2019notsofast}. WebAssembly speeds up single-thread execution and addresses a different concern than ParaWeb's Worker-Thread parallelism, so the two combine naturally by running WebAssembly-compiled functions inside the worker pool. WebGPU~\cite{w3c-webgpu} is a more recent W3C standard that provides low-level access to GPU compute and rendering capabilities from JavaScript. Unlike its predecessor WebGL, which was designed for graphics rendering, WebGPU includes a compute shader pipeline that allows general-purpose GPU (GPGPU) computations using the WGSL shading language. WebGPU is available in Node.js through packages such as \texttt{webgpu} (Dawn-based). Recent work has begun characterising its performance for general-purpose scientific computing in the browser~\cite{aldahir2022webgpu}. GPU.js~\cite{sapuan2018gpujs} is an earlier JavaScript library that transpiles JavaScript functions to WebGL shader code for GPU execution, but it offers only Map-like operations and does not support the full range of parallel patterns.

In contrast, ParaWeb is a pure user-space library built on standardized APIs (Node.js Worker Threads, browser Web Workers, and WebGPU compute shaders), and therefore deploys to any modern runtime today without engine changes. It also covers a broader pattern set than River Trail, adding Stencil, MapReduce, and the task-parallel Farm, Pipeline, and Divide-and-Conquer patterns, as well as stream stage operators (filter, window-reduce, iterate) inside Pipeline.
\section{Design}\label{sec:design}

This section describes the architecture of ParaWeb and the design decisions behind its three implementation variants. We first present the overall architecture and the rationale for restricting input data to numeric arrays, then classify the ten supported patterns by their computational characteristics, and finally describe the trade-offs between the message-passing, shared-buffer, and GPU execution backends.

\subsection{Architecture Overview}

ParaWeb follows a modular architecture in which each parallel pattern is implemented as a separate, self-contained class. This design promotes code reuse, simplifies testing and maintenance, and allows individual patterns to evolve independently. Despite their independence, all patterns share a common set of design principles that ensure consistency across the library.

Each pattern exposes a unified interface with an optional \texttt{numThreads} parameter that allows developers to control the degree of parallelism. When the argument is omitted, ParaWeb defaults to the number of available processing cores, queried via \texttt{os.cpus().length} in Node.js or \texttt{navigator.hardwareConcurrency} in the browser, so applications use all available cores without manual configuration. The patterns handle all Worker Thread management details, including thread creation, task distribution, result collection, and resource cleanup. TypeScript provides compile-time type checking, so that type errors are caught during development rather than at runtime. Error handling ensures that worker failures are properly managed, with resource cleanup even under error conditions.

\subsection{Data Type Restriction}

ParaWeb restricts input data to numeric arrays. The MP variants support one-dimensional and two-dimensional numeric arrays, while the Shared variants operate on one-dimensional numeric arrays backed by \texttt{Float64Array} buffers. This design decision is motivated by several considerations. First, it simplifies the implementation by removing complex type detection and polymorphic handling logic. Second, it improves performance by avoiding type checking overhead on the parallel execution path. Third, it aligns with established frameworks such as GrPPI~\cite{grppi} and SkePU~\cite{skepu3}, which similarly focus on structured numeric data. Fourth, the JavaScript engine can make stronger assumptions about data layout and memory access patterns when operating on typed arrays, which allows for better optimization.

This restriction is justified by the observation that most computationally intensive parallel workloads operate on numeric data. Application domains such as image processing, signal analysis, scientific computation, financial modeling, and data transformation pipelines mostly work with numeric arrays. By focusing on this common case, ParaWeb achieves simplicity and performance without sacrificing expressiveness for the workloads that benefit most from parallelization.

\subsection{Pattern Classification}

The ten patterns implemented in ParaWeb can be classified into three categories based on their computational characteristics. Data-parallel patterns, including Map, Filter, Reduce, Scan, and Scatter, apply a uniform operation across a collection of elements. Their inter-element dependencies are either absent (Map, Filter) or structurally regular (associative in Reduce and Scan, index-driven in Scatter), which admits a partition-compute-combine parallel decomposition. Task-parallel patterns, including Farm, Pipeline, and Divide-and-Conquer, focus on organizing work into tasks that can be distributed across workers, with varying degrees of dependency and coordination between tasks. Specialized patterns, including Stencil and MapReduce, address specific computational structures that arise frequently in practice but require specialized handling, such as neighborhood-based computation or combined transformation and aggregation.

Figure~\ref{fig:pattern_overview} provides abstract views of each pattern, illustrating the data flow from inputs to outputs. Detailed descriptions of how each pattern is implemented in ParaWeb are presented later in Section~\ref{sec:implementation}.

\tikzset{
  pbox/.style={draw,rounded corners,align=center,inner sep=2pt,minimum height=5mm},
  pnode/.style={draw,circle,inner sep=1.2pt,minimum size=3.6mm},
  parrow/.style={-Latex,thick},
  pdash/.style={draw,dashed,rounded corners,inner sep=2pt}
}

\begin{figure*}[!htbp]
\centering
\scriptsize

\begin{minipage}[t]{0.32\textwidth}
\centering
\begin{tikzpicture}[node distance=2mm]
\node (i1) {$x_1$};
\node[right=2mm of i1] (i2) {$x_2$};
\node[right=2mm of i2] (i3) {$x_3$};
\node[right=2mm of i3] (i4) {$x_4$};
\node[below=3mm of i1] (f0) [pnode] {$f$};
\node[below=3mm of i2] (f1) [pnode] {$f$};
\node[below=3mm of i3] (f2) [pnode] {$f$};
\node[below=3mm of i4] (f3) [pnode] {$f$};
\node[below=3mm of f0] (o1) {$y_1$};
\node[below=3mm of f1] (o2) {$y_2$};
\node[below=3mm of f2] (o3) {$y_3$};
\node[below=3mm of f3] (o4) {$y_4$};
\draw[parrow] (i1) -- (f0);
\draw[parrow] (i2) -- (f1);
\draw[parrow] (i3) -- (f2);
\draw[parrow] (i4) -- (f3);
\draw[parrow] (f0) -- (o1);
\draw[parrow] (f1) -- (o2);
\draw[parrow] (f2) -- (o3);
\draw[parrow] (f3) -- (o4);
\end{tikzpicture}
\par\vspace{1mm}\textbf{Map}
\end{minipage}%
\hfill
\begin{minipage}[t]{0.32\textwidth}
\centering
\begin{tikzpicture}[node distance=2mm]
\node (i1) {$x_1$};
\node[right=2mm of i1] (i2) {$x_2$};
\node[right=2mm of i2] (i3) {$x_3$};
\node[right=2mm of i3] (i4) {$x_4$};
\node[below=3mm of i1] (p0) [pnode] {$p$};
\node[below=3mm of i2] (p1) [pnode] {$p$};
\node[below=3mm of i3] (p2) [pnode] {$p$};
\node[below=3mm of i4] (p3) [pnode] {$p$};
\node[below=3mm of p0] (o1) {$x_1$};
\node[below=3mm of p2] (o3) {$x_3$};
\draw[parrow] (i1) -- (p0);
\draw[parrow] (i2) -- (p1);
\draw[parrow] (i3) -- (p2);
\draw[parrow] (i4) -- (p3);
\draw[parrow] (p0) -- (o1);
\draw[parrow,opacity=0.3] (p1) -- ++(0,-4mm);
\draw[parrow] (p2) -- (o3);
\draw[parrow,opacity=0.3] (p3) -- ++(0,-4mm);
\end{tikzpicture}
\par\vspace{1mm}\textbf{Filter}
\end{minipage}%
\hfill
\begin{minipage}[t]{0.32\textwidth}
\centering
\begin{tikzpicture}[node distance=2mm]
\node (i1) {$x_1$};
\node[right=2mm of i1] (i2) {$x_2$};
\node[right=2mm of i2] (i3) {$x_3$};
\node[right=2mm of i3] (i4) {$x_4$};
\node[below=3mm of i1] (s1) [pnode] {1};
\node[below=3mm of i2] (s2) [pnode] {3};
\node[below=3mm of i3] (s3) [pnode] {0};
\node[below=3mm of i4] (s4) [pnode] {2};
\node[below=4mm of s1] (o1) {$y_1$};
\node[below=4mm of s2] (o3) {$y_3$};
\node[below=4mm of s3] (o0) {$y_0$};
\node[below=4mm of s4] (o2) {$y_2$};
\draw[parrow] (i1) -- (s1);
\draw[parrow] (i2) -- (s2);
\draw[parrow] (i3) -- (s3);
\draw[parrow] (i4) -- (s4);
\draw[parrow] (s1) -- (o1);
\draw[parrow] (s2) -- (o3);
\draw[parrow] (s3) -- (o0);
\draw[parrow] (s4) -- (o2);
\end{tikzpicture}
\par\vspace{1mm}\textbf{Scatter}
\end{minipage}

\vspace{3mm}

\begin{minipage}[t]{0.32\textwidth}
\centering
\begin{tikzpicture}[node distance=2mm]
\node (i1) {$x_1$};
\node[right=2mm of i1] (i2) {$x_2$};
\node[right=2mm of i2] (i3) {$x_3$};
\node[right=2mm of i3] (i4) {$x_4$};
\node[below=4mm of i2] (r1) [pnode] {$\oplus$};
\node[below=4mm of i3] (r2) [pnode] {$\oplus$};
\node[below=4mm of r1] (r3) [pnode] {$\oplus$};
\node[below=4mm of r3] (out) {$y$};
\draw[parrow] (i1) -- (r1);
\draw[parrow] (i2) -- (r1);
\draw[parrow] (i3) -- (r2);
\draw[parrow] (i4) -- (r2);
\draw[parrow] (r1) -- (r3);
\draw[parrow] (r2) -- (r3);
\draw[parrow] (r3) -- (out);
\end{tikzpicture}
\par\vspace{1mm}\textbf{Reduce}
\end{minipage}%
\hfill
\begin{minipage}[t]{0.32\textwidth}
\centering
\begin{tikzpicture}[node distance=2mm]
\node (i1) {$x_1$};
\node[right=2mm of i1] (i2) {$x_2$};
\node[right=2mm of i2] (i3) {$x_3$};
\node[right=2mm of i3] (i4) {$x_4$};
\node[below=4mm of i1] (s1) [pnode] {$\oplus$};
\node[below=4mm of i2] (s2) [pnode] {$\oplus$};
\node[below=4mm of i3] (s3) [pnode] {$\oplus$};
\node[below=4mm of i4] (s4) [pnode] {$\oplus$};
\node[below=3mm of s1] (o1) {$y_1$};
\node[below=3mm of s2] (o2) {$y_2$};
\node[below=3mm of s3] (o3) {$y_3$};
\node[below=3mm of s4] (o4) {$y_4$};
\draw[parrow] (i1) -- (s1);
\draw[parrow] (i2) -- (s2);
\draw[parrow] (i3) -- (s3);
\draw[parrow] (i4) -- (s4);
\draw[parrow] (s1) -- (s2);
\draw[parrow] (s2) -- (s3);
\draw[parrow] (s3) -- (s4);
\draw[parrow] (s1) -- (o1);
\draw[parrow] (s2) -- (o2);
\draw[parrow] (s3) -- (o3);
\draw[parrow] (s4) -- (o4);
\end{tikzpicture}
\par\vspace{1mm}\textbf{Scan}
\end{minipage}%
\hfill
\begin{minipage}[t]{0.32\textwidth}
\centering
\begin{tikzpicture}[node distance=2mm]
\node (i1) {$x_1$};
\node[right=2mm of i1] (i2) {$x_2$};
\node[right=2mm of i2] (i3) {$x_3$};
\node[right=2mm of i3] (i4) {$x_4$};
\node[right=2mm of i4] (i5) {$x_5$};
\node[below=4mm of i2] (s2) [pnode] {$f$};
\node[below=4mm of i3] (s3) [pnode] {$f$};
\node[below=4mm of i4] (s4) [pnode] {$f$};
\node[below=3mm of s2] (o2) {$y_2$};
\node[below=3mm of s3] (o3) {$y_3$};
\node[below=3mm of s4] (o4) {$y_4$};
\draw[parrow] (i1) -- (s2);
\draw[parrow] (i2) -- (s2);
\draw[parrow] (i3) -- (s2);
\draw[parrow] (i2) -- (s3);
\draw[parrow] (i3) -- (s3);
\draw[parrow] (i4) -- (s3);
\draw[parrow] (i3) -- (s4);
\draw[parrow] (i4) -- (s4);
\draw[parrow] (i5) -- (s4);
\draw[parrow] (s2) -- (o2);
\draw[parrow] (s3) -- (o3);
\draw[parrow] (s4) -- (o4);
\end{tikzpicture}
\par\vspace{1mm}\textbf{Stencil}
\end{minipage}

\vspace{3mm}


\begin{minipage}[t]{0.24\textwidth}
\centering
\begin{tikzpicture}[node distance=2mm]
\node (in) {in};
\node[pnode, below=2mm of in] (s1) {$f_1$};
\node[pnode, below=2mm of s1] (s2) {$f_2$};
\node[pnode, below=2mm of s2] (s3) {$f_n$};
\node[below=2mm of s3] (out) {out};
\draw[parrow] (in) -- (s1);
\draw[parrow] (s1) -- (s2);
\draw[parrow] (s2) -- (s3);
\draw[parrow] (s3) -- (out);
\end{tikzpicture}
\par\vspace{1mm}\textbf{Pipeline}
\end{minipage}%
\hfill
\begin{minipage}[t]{0.24\textwidth}
\centering
\begin{tikzpicture}[node distance=2mm]
\node (i1) {$x_1$};
\node[right=1mm of i1] (i2) {$x_2$};
\node[right=1mm of i2] (i3) {$x_3$};
\node[right=1mm of i3] (i4) {$x_4$};
\node[below=2mm of i1] (f1) [pnode] {$f$};
\node[below=2mm of i2] (f2) [pnode] {$f$};
\node[below=2mm of i3] (f3) [pnode] {$f$};
\node[below=2mm of i4] (f4) [pnode] {$f$};
\draw[parrow] (i1) -- (f1);
\draw[parrow] (i2) -- (f2);
\draw[parrow] (i3) -- (f3);
\draw[parrow] (i4) -- (f4);
\node[below=3mm of f1.south east, xshift=1mm] (r1) [pnode] {$\oplus$};
\node[below=3mm of f3.south east, xshift=1mm] (r2) [pnode] {$\oplus$};
\node[below=3mm of $(r1)!0.5!(r2)$] (r3) [pnode] {$\oplus$};
\node[below=2mm of r3] (out) {$y$};
\draw[parrow] (f1) -- (r1);
\draw[parrow] (f2) -- (r1);
\draw[parrow] (f3) -- (r2);
\draw[parrow] (f4) -- (r2);
\draw[parrow] (r1) -- (r3);
\draw[parrow] (r2) -- (r3);
\draw[parrow] (r3) -- (out);
\end{tikzpicture}
\par\vspace{1mm}\textbf{MapReduce}
\end{minipage}%
\hfill
\begin{minipage}[t]{0.24\textwidth}
\centering
\begin{tikzpicture}[node distance=2mm]
\node (t1) {$t_1$};
\node[right=1.2mm of t1] (t2) {$t_2$};
\node[right=1.2mm of t2] (t3) {$t_3$};
\node[right=1.2mm of t3] (tdots) {$\cdots$};
\node[pnode, below=3mm of t2.south east, xshift=2mm] (e) {E};
\node[pnode, below=3mm of e, xshift=-7mm] (w1) {W};
\node[pnode, below=3mm of e] (w2) {W};
\node[pnode, below=3mm of e, xshift=7mm] (w3) {W};
\node[right=1.2mm of w3] (wdots) {$\cdots$};
\node[below=3mm of w2] (r1) {$r_1$};
\node[left=1.2mm of r1] (r2) {$r_2$};
\node[right=1.2mm of r1] (r3) {$r_3$};
\node[right=1.2mm of r3] (rdots) {$\cdots$};
\draw[parrow] (t1) -- (e);
\draw[parrow] (t2) -- (e);
\draw[parrow] (t3) -- (e);
\draw[parrow] (tdots) -- (e);
\draw[parrow] (e) -- (w1);
\draw[parrow] (e) -- (w2);
\draw[parrow] (e) -- (w3);
\draw[parrow] (e) -- (wdots);
\draw[parrow] (w1) -- (r2);
\draw[parrow] (w2) -- (r1);
\draw[parrow] (w3) -- (r3);
\draw[parrow] (wdots) -- (rdots);
\end{tikzpicture}
\par\vspace{1mm}\textbf{Farm}
\end{minipage}%
\hfill
\begin{minipage}[t]{0.24\textwidth}
\centering
\begin{tikzpicture}[node distance=2mm]
\node (in) {in};
\node[pbox, below=2mm of in] (split) {split};
\node[pbox, below=4mm of split, xshift=-7mm] (sub1) {sub};
\node[pbox, below=4mm of split, xshift=7mm] (sub2) {sub};
\node[pbox, below=4mm of split, yshift=-10mm] (comb) {combine};
\node[below=2mm of comb] (out) {out};
\draw[parrow] (in) -- (split);
\draw[parrow] (split) -- (sub1);
\draw[parrow] (split) -- (sub2);
\draw[parrow] (sub1) -- (comb);
\draw[parrow] (sub2) -- (comb);
\draw[parrow] (comb) -- (out);
\draw[->,dashed,thick,gray]
  (sub1.west) .. controls +(-3mm,3mm) and +(-5mm,-3mm) .. (split.west);
\end{tikzpicture}
\par\vspace{1mm}\textbf{D\&C}
\end{minipage}

\caption{Abstract views of the ten ParaWeb patterns.}
\label{fig:pattern_overview}
\end{figure*}
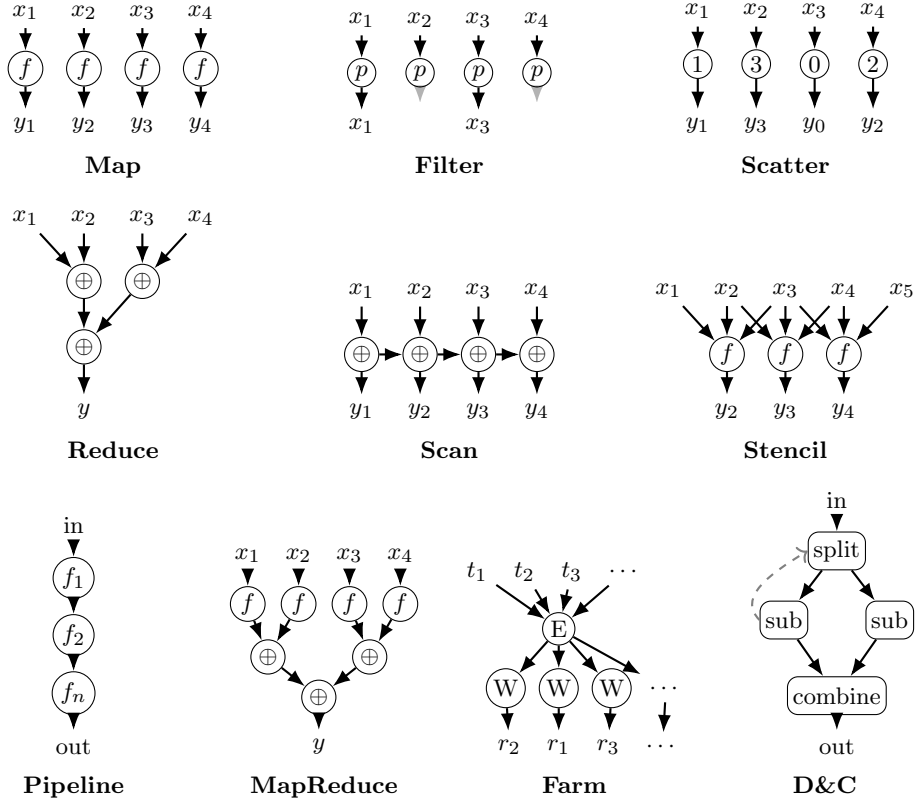

\subsection{Three Implementation Variants}\label{subsec:variants}

For each pattern, ParaWeb provides three implementation variants that explore different points in the parallelism trade-off space: a message-passing (MP) variant prioritizing simplicity and broad data-type support at the cost of serialization overhead; a shared-buffer (Shared) variant that eliminates copying for numeric data at the cost of explicit coordination and a typed-array restriction; and a GPU variant offering the highest throughput for compute-bound data-parallel patterns but requiring WGSL operations and \texttt{f32} precision. Section~\ref{sec:implementation} describes each variant in detail.

\subsection{Programming Interface}\label{subsec:api}

ParaWeb exposes each pattern as a free function under a single \texttt{paraweb} namespace, so users write parallel code that looks like ordinary sequential JavaScript. An equivalent Map implementation using raw Node.js Worker Threads typically spans roughly 40 lines across two files: a main script that chunks the input, spawns workers, wires \texttt{message}, \texttt{error}, and \texttt{exit} handlers, awaits all promises, and concatenates partial results; and a separate worker script that rebuilds the user function from its source string and posts the per-chunk result. With ParaWeb the same workload is one line. Listing~\ref{lst:api} lists the public interface of all ten patterns, and the three variants (MP, Shared, GPU) are reached through \texttt{paraweb.mp.*}, \texttt{paraweb.shared.*}, and \texttt{paraweb.gpu.*}; the top-level names (\texttt{paraweb.map}, \texttt{paraweb.reduce}, $\dots$) select the empirically best variant per pattern (Section~\ref{sec:evaluation}).

\begin{lstlisting}[caption={Public interface of the ten ParaWeb patterns. Each call returns a \texttt{Promise}; \texttt{threads} defaults to the number of available CPU cores.}, label={lst:api}]
import paraweb from "paraweb";

// Data-parallel
paraweb.map(fn, input, threads?);
paraweb.filter(pred, input, threads?);
paraweb.reduce(op, input, identity, threads?);
paraweb.scan(op, input, identity, threads?);
paraweb.scatter(input, indices, default?, conflictFn?, threads?);

// Task-parallel
paraweb.farm(fn, input, threads?);
paraweb.pipeline(stages, input, threads?);
paraweb.divideAndConquer(divideFn, conquerFn, baseFn, input, threads?);

// Specialized
paraweb.stencil(fn, input, window, threads?, edgeOption?);
paraweb.mapReduce(mapFn, reduceOp, input, threads?);
\end{lstlisting}

The three variants share the same calling convention, so switching between them is a single property access. Listing~\ref{lst:variants} shows the same Map across MP, Shared, and GPU. The GPU variant accepts a WGSL expression (or a built-in operation name) instead of a JavaScript function, because the body executes on the device. The call site is otherwise identical.

\begin{lstlisting}[caption={Switching between the three implementation variants.}, label={lst:variants}]
import paraweb from "paraweb";
const f = (x: number) => Math.sin(x) * Math.cos(x * 0.5);
const y1 = await paraweb.mp.map(f, input);      // Worker Threads, postMessage
const y2 = await paraweb.shared.map(f, input);  // SharedArrayBuffer, zero-copy
const y3 = await paraweb.gpu.map(               // WebGPU compute shader
  { wgsl: "sin(x) * cos(x * 0.5)" }, input);
\end{lstlisting}

Composition of patterns is expressed through ordinary JavaScript control flow. Listing~\ref{lst:pipeline} builds a three-stage processing pipeline (denoise, feature extraction, classification). Since every pattern returns a \texttt{Promise}, patterns compose naturally with \texttt{await} and can be mixed with non-parallel code without any special syntax.

\begin{lstlisting}[caption={Three-stage image-processing pipeline.}, label={lst:pipeline}]
import paraweb from "paraweb";
const denoise  = (x: number) => 0.25 * x + 0.5 * x + 0.25 * x;
const features = (x: number) => Math.tanh(x);
const classify = (x: number) => x > 0.5 ? 1 : 0;
const labels = await paraweb.pipeline(
  [denoise, features, classify], pixels, /* threads */ 16);
\end{lstlisting}

\section{Implementation}\label{sec:implementation}

This section describes the implementation of ParaWeb, including the worker thread management infrastructure shared across patterns, the function serialization and communication protocol used between the main thread and workers, the implementation of the three execution variants (MP, Shared, GPU), and the pattern-specific implementation details for each of the ten patterns.

\subsection{Worker Thread Management}

All worker-based patterns in ParaWeb use a reusable worker pool (\texttt{WorkerPool}) as the underlying mechanism for parallel execution, using Node.js Worker Threads on the server and Web Workers in the browser. Workers are lazily created, cached by script path and pool size, and reused across invocations to amortize creation overhead, an approach broadly consistent with prior work on dynamic Web Worker pool management for JavaScript applications~\cite{verdu2016pool}. The input is partitioned into $\lceil n/p \rceil$-element chunks, where $n$ is the array length and $p$ the number of threads. The thread count $p$ is capped at $\min(p, n)$ to avoid spawning empty workers on small inputs.

\subsection{Function Serialization and Communication}

Worker Threads and Web Workers operate in separate memory spaces, which means that functions cannot be passed directly between threads. ParaWeb addresses this through function serialization: user-supplied functions are converted to strings using \texttt{Function.toString()} and reconstructed in worker threads using \texttt{new Function()}, which compiles the function in the worker's global scope. Note that \texttt{new Function()} does not preserve the surrounding lexical scope, so the user-supplied function must be self-contained: references to external helper functions or to captured variables from the enclosing scope are not resolved in the worker and will throw at execution time. In practice this means user code either inlines the helper logic inside the function body, or passes additional data as part of the input array rather than as a closed-over variable. Built-in globals such as \texttt{Math}, \texttt{Array}, and typed-array constructors are available in the worker and can be used normally. Task messages in MP variants contain the serialized function code, input data, and pattern-specific parameters, while Shared variants carry buffer references, index ranges, and coordination metadata such as atomic counters. Workers report errors using an \texttt{"error"} string protocol and shut down upon receiving a \texttt{"terminate"} message.

\subsection{MP and Shared CPU Variants}

The MP variants partition input arrays into chunks and transfer data to workers using \texttt{postMessage()}, which relies on the structured clone algorithm and incurs copying overhead proportional to input size. The Shared variants place input and output data in \texttt{SharedArrayBuffer} instances backed by \texttt{Float64Array} views, allowing workers to read and write shared memory regions without copying. The generic \texttt{divideAndConquer} API still delegates to MP-style transfer, since arbitrary user-supplied subproblems are not guaranteed to be array-shaped. The specialised Shared D\&C paths used in our evaluation (\texttt{fft} and \texttt{sort}) operate directly on shared buffers, with workers dispatched per Cooley-Tukey stage. While shared buffers eliminate transfer overhead for large arrays, they introduce synchronization and coordination costs. For instance, the Shared Filter variant uses a two-pass approach (count matching elements, then write results), and the Shared Farm variant uses an atomic counter to distribute work dynamically. Our experimental results show that the Shared variant outperforms MP for most patterns on large datasets.

\subsection{GPU Variants via WebGPU}

The GPU variants offload computation to the graphics processor using WebGPU compute shaders. They accept operations specified as WGSL expressions or predefined built-in operation names, such as \texttt{"square"}, \texttt{"collatz\_steps"}, or \texttt{"is\_prime"}, which are injected into a compute shader template. The GPU infrastructure consists of four core modules: (i) a singleton GPU device manager that acquires and caches the \texttt{GPUDevice}, (ii) buffer utilities for uploading \texttt{Float32Array} data to GPU storage buffers and reading back results, (iii) a pipeline cache that avoids shader recompilation across repeated invocations, and (iv) a shader builder that generates WGSL compute shaders from operation specifications. Note that WebGPU compute shaders only support 32-bit floating-point arithmetic (\texttt{f32}), so GPU results have lower numerical precision than the \texttt{Float64Array}-based CPU variants. This trade-off is acceptable for workloads where throughput matters more than precision, such as image processing, scientific visualization, and data analytics.

For data-parallel patterns such as Map, Filter, and Stencil, each input element maps directly to a GPU thread (invocation), with a workgroup size of 256 threads. The GPU hardware scheduler distributes workgroups across compute units, providing automatic load balancing. For Reduce, we use a hierarchical reduction scheme with workgroup-level parallel reduction using shared memory, followed by iterative passes until a single value remains. Filter uses a three-phase stream compaction approach: a mark phase that evaluates the predicate, a prefix-sum phase that computes output indices, and a compact phase that writes matching elements to the output buffer. For task-parallel patterns such as Farm and Pipeline, the GPU variants delegate to the Map kernel, since the GPU's hardware scheduler provides work distribution analogous to the CPU Farm pattern.

\subsection{Pattern-Specific Implementations}

The five \emph{data-parallel} patterns all start by chunking the input into $p$ aligned slices. Map and Filter apply the user function or predicate independently per chunk and concatenate the results, with Filter's Shared variant using a two-pass count-then-write approach to handle variable-length output. Scatter redistributes values to destinations given by an index array, with each worker producing index-value pairs and the main thread assembling an output of length $\max(\textit{index}) + 1$, resolving collisions with a last-write policy or optional combiner (matching River Trail's scatter semantics~\cite{herhut2013rivertrail}). Reduce aggregates each chunk to a partial result in parallel and combines the partials on the main thread using the same associative operator, with the user-supplied \texttt{identity} value as the initial accumulator. Scan (inclusive prefix-scan) uses a two-pass parallel algorithm: each worker computes a local inclusive scan and reports its chunk total, the main thread computes an exclusive scan of chunk totals, and each worker then adds its prefix offset to every element in its chunk. Reduce and Scan both require an associative binary operation, with $O(n/p + p)$ complexity.

The three \emph{task-parallel} patterns add coordination on top of the worker pool. Farm distributes variable-cost tasks across workers using one of two strategies depending on the variant. The MP variant uses static chunking ($p$ contiguous ranges, one per worker), since per-task message passing would dominate cost. The Shared variant uses a lock-free atomic counter so that workers self-schedule by pulling batches of $\lceil n/p \rceil$ indices via \texttt{Atomics.add} on a shared counter, avoiding both per-task messages and main-thread coordination. Pipeline processes data through sequential stages with intra-stage data parallelism, where each stage is either a plain function (applied element-wise as a parallel map) or one of three stream stage operators inspired by GrPPI's streaming patterns~\cite{grppi}: \emph{stream filter} (drops items failing a predicate), \emph{stream window-reduce} (collapses each sliding window into one value), and \emph{stream iterate} (applies a fixed-point loop per item). Divide-and-Conquer (D\&C) recursively decomposes problems through a user-supplied \texttt{divideFn}, distributes top-level subproblems across threads, and combines results via \texttt{conquerFn}. The base-case threshold trades more parallelism for more task-management overhead.

The two \emph{specialized} patterns handle recurring computational structures. Stencil applies neighborhood-based computation with overlap regions of $\lfloor s/2 \rfloor$ elements around each worker's slice, so boundary computations resolve without inter-worker communication. Configurable edge handling makes it suitable for image convolution, signal processing, and finite-difference methods. MapReduce composes a parallel Map phase with a subsequent Reduce phase by reusing both implementations, mirroring the computational structure of Hadoop and Spark adapted for shared-memory worker threads.

\section{Experimental Setup}\label{sec:setup}

To evaluate the performance of ParaWeb across different workloads, data sizes, and thread configurations, we designed a set of benchmarks that combine representative data sizes with computationally intensive per-pattern algorithms. In this section, we describe the test configurations and the specific algorithms used to exercise each pattern.

\subsection{Test Configurations}

We evaluate each pattern across three data sizes of increasing scale: Medium (100K elements), Large (1M elements), and Extremely Large (5M--10M elements). For each data size, we test a plain-JavaScript sequential baseline plus four parallel configurations of 2, 4, 8, and 16 threads. These size configurations correspond to representative use cases: data transformation pipelines (100K), scientific computation (1M), and large-scale data analytics (5M--10M).

Each configuration is executed for five runs after two warmup runs, and we report the mean execution time. Speedup is the ratio of a plain-JavaScript sequential baseline (main thread, no workers, no \texttt{SharedArrayBuffer}) to the parallel time, using identical inputs and the same per-element workload as the parallel run.

\subsection{Test Algorithms}

Each pattern is evaluated using a computationally intensive algorithm designed to produce sufficient per-element work to amortize parallelization overhead:

\begin{itemize}
    \item \textbf{Map}: A 50-iteration iterative transform per element (\(\sin\), \(\cos\), \(\sqrt{\,}\), and a polynomial step), providing a compute-bound workload with no inter-element dependencies.
    \item \textbf{Filter}: A 30-iteration trigonometric predicate followed by a threshold comparison, which produces data-dependent output size with a roughly 50\% pass rate on the chosen input distribution.
    \item \textbf{Reduce}: Associative sum over a 10-iteration sin/cos per-element transform (roughly 1/5 the arithmetic intensity of Map/MapReduce). The transform is supplied as a separate \texttt{mapFn} argument so the combine operator stays associative.
    \item \textbf{Scan}: Inclusive prefix sum (associative) fused with the 50-iteration trigonometric transform per element, so that the two-pass prefix-sum algorithm is exercised on a compute-bound workload with enough per-element work to amortize the phase-2 offset pass.
    \item \textbf{Scatter}: Index remapping with randomly generated destination indices in \([0, n/4)\), applying the 50-iteration trigonometric transform per element (matching Map's arithmetic intensity) and summing values at conflicting positions on CPU.
    \item \textbf{Stencil}: Weighted five-point neighborhood sum (weights \([1, 2, 3, 2, 1]\)) followed by a 15-iteration trigonometric refinement per element, producing high arithmetic intensity per stencil application.
    \item \textbf{Farm}: Per-task workload over starting numbers in \([1000, 2000)\) combining a Collatz step-count (variable cost, $\sim$ 50--250 steps) with a 200-iteration trigonometric refinement. The Collatz prefix preserves Farm's variable-cost characteristic while the refinement raises per-task cost into the millisecond range so coordination amortizes.
    \item \textbf{Divide-and-Conquer}: Parallel Cooley-Tukey radix-2 FFT~\cite{cooley1965fft} on a complex array of $N$ points.
    \item \textbf{Pipeline}: Two stages, a 30-iteration trigonometric transform followed by a 20-iteration polynomial evaluation with rounding, testing sequential stage composition with intra-stage parallelism.
    \item \textbf{MapReduce}: The 50-iteration trigonometric transform used by Map, followed by associative sum aggregation, testing the fused map-then-reduce workflow.
\end{itemize}

The same per-element transforms are fused into the WGSL kernels for Reduce, Scan, and Scatter so that GPU and CPU measurements in Table~\ref{tab:gpu_comparison} compare equivalent workloads.

\section{Experimental Evaluation}\label{sec:evaluation}


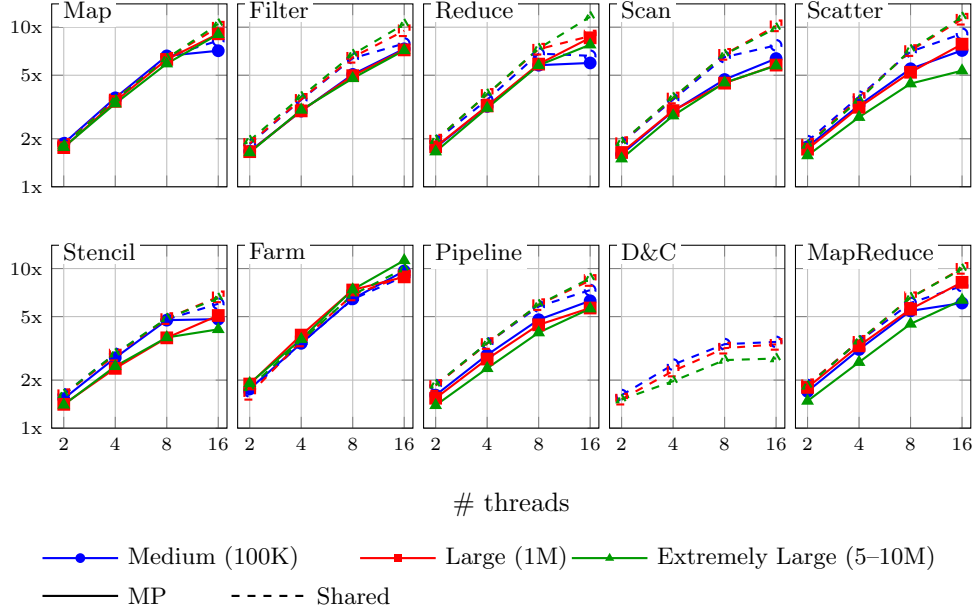
\begin{figure*}[!htbp]
\centering
\begin{tikzpicture}
\begin{groupplot}[
  group style={
    group size=5 by 2,
    horizontal sep=4pt,
    vertical sep=22pt,
    xticklabels at=edge bottom,
    yticklabels at=edge left,
  },
  width=3.9cm,
  height=4.0cm,
  ymode=log,
  log basis y=10,
  ytick={1,2,5,10},
  yticklabels={1x,2x,5x,10x},
  ymin=1, ymax=14,
  xmode=log,
  log basis x=2,
  xtick={2,4,8,16},
  xticklabels={2,4,8,16},
  xmin=1.7, xmax=18,
  grid=major,
  title style={at={(axis description cs:0.04, 0.95)}, anchor=north west, font=\scriptsize, fill=white, inner sep=1.5pt},
  tick label style={font=\footnotesize},
  label style={font=\footnotesize},
  every axis y label/.style={at={(axis description cs:-0.22, 0.5)}, rotate=90, anchor=south},
  every axis x label/.style={at={(axis description cs:0.5, -0.30)}, anchor=north},
]
\nextgroupplot[title={Map}, ymin=1, ytick={1,2,5,10}, yticklabels={1x,2x,5x,10x}]
\addplot[color=blue,mark=*,thick] coordinates { (2,1.87) (4,3.61) (8,6.61) (16,7.12) };

\addplot[color=blue,mark=o,thick,dashed] coordinates { (2,1.88) (4,3.50) (8,6.54) (16,8.23) };

\addplot[color=red,mark=square*,thick] coordinates { (2,1.76) (4,3.43) (8,6.28) (16,9.09) };

\addplot[color=red,mark=square,thick,dashed] coordinates { (2,1.84) (4,3.55) (8,6.38) (16,10.00) };

\addplot[color=green!60!black,mark=triangle*,thick] coordinates { (2,1.77) (4,3.32) (8,5.91) (16,9.03) };

\addplot[color=green!60!black,mark=triangle,thick,dashed] coordinates { (2,1.83) (4,3.51) (8,6.37) (16,10.48) };

\nextgroupplot[title={Filter}, ymin=1, ytick={1,2,5,10}, yticklabels={}]
\addplot[color=blue,mark=*,thick] coordinates { (2,1.67) (4,2.98) (8,5.06) (16,7.32) };

\addplot[color=blue,mark=o,thick,dashed] coordinates { (2,1.85) (4,3.51) (8,6.39) (16,7.90) };

\addplot[color=red,mark=square*,thick] coordinates { (2,1.66) (4,2.99) (8,4.97) (16,7.21) };

\addplot[color=red,mark=square,thick,dashed] coordinates { (2,1.87) (4,3.53) (8,6.48) (16,9.50) };

\addplot[color=green!60!black,mark=triangle*,thick] coordinates { (2,1.64) (4,3.03) (8,4.78) (16,7.16) };

\addplot[color=green!60!black,mark=triangle,thick,dashed] coordinates { (2,1.95) (4,3.67) (8,6.75) (16,10.44) };

\nextgroupplot[title={Reduce}, ymin=1, ytick={1,2,5,10}, yticklabels={}]
\addplot[color=blue,mark=*,thick] coordinates { (2,1.76) (4,3.19) (8,5.78) (16,5.97) };

\addplot[color=blue,mark=o,thick,dashed] coordinates { (2,1.94) (4,3.51) (8,6.83) (16,6.62) };

\addplot[color=red,mark=square*,thick] coordinates { (2,1.78) (4,3.21) (8,5.87) (16,8.58) };

\addplot[color=red,mark=square,thick,dashed] coordinates { (2,1.96) (4,3.79) (8,7.28) (16,8.80) };

\addplot[color=green!60!black,mark=triangle*,thick] coordinates { (2,1.66) (4,3.11) (8,5.77) (16,7.77) };

\addplot[color=green!60!black,mark=triangle,thick,dashed] coordinates { (2,1.96) (4,3.81) (8,7.35) (16,11.62) };

\nextgroupplot[title={Scan}, ymin=1, ytick={1,2,5,10}, yticklabels={}]
\addplot[color=blue,mark=*,thick] coordinates { (2,1.60) (4,2.98) (8,4.70) (16,6.36) };

\addplot[color=blue,mark=o,thick,dashed] coordinates { (2,1.88) (4,3.52) (8,6.49) (16,7.70) };

\addplot[color=red,mark=square*,thick] coordinates { (2,1.64) (4,3.00) (8,4.46) (16,5.79) };

\addplot[color=red,mark=square,thick,dashed] coordinates { (2,1.89) (4,3.60) (8,6.76) (16,10.20) };

\addplot[color=green!60!black,mark=triangle*,thick] coordinates { (2,1.50) (4,2.79) (8,4.49) (16,5.74) };

\addplot[color=green!60!black,mark=triangle,thick,dashed] coordinates { (2,1.90) (4,3.62) (8,6.74) (16,9.88) };

\nextgroupplot[title={Scatter}, ymin=1, ytick={1,2,5,10}, yticklabels={}]
\addplot[color=blue,mark=*,thick] coordinates { (2,1.78) (4,3.24) (8,5.49) (16,7.15) };

\addplot[color=blue,mark=o,thick,dashed] coordinates { (2,1.93) (4,3.67) (8,7.03) (16,9.09) };

\addplot[color=red,mark=square*,thick] coordinates { (2,1.72) (4,3.15) (8,5.22) (16,7.83) };

\addplot[color=red,mark=square,thick,dashed] coordinates { (2,1.85) (4,3.57) (8,7.14) (16,11.22) };

\addplot[color=green!60!black,mark=triangle*,thick] coordinates { (2,1.57) (4,2.73) (8,4.43) (16,5.34) };

\addplot[color=green!60!black,mark=triangle,thick,dashed] coordinates { (2,1.83) (4,3.50) (8,7.23) (16,11.54) };

\nextgroupplot[title={Stencil}, ymin=1, ytick={1,2,5,10}, yticklabels={1x,2x,5x,10x}]
\addplot[color=blue,mark=*,thick] coordinates { (2,1.53) (4,2.76) (8,4.75) (16,4.82) };

\addplot[color=blue,mark=o,thick,dashed] coordinates { (2,1.63) (4,2.92) (8,4.80) (16,6.02) };

\addplot[color=red,mark=square*,thick] coordinates { (2,1.41) (4,2.37) (8,3.68) (16,5.08) };

\addplot[color=red,mark=square,thick,dashed] coordinates { (2,1.62) (4,2.87) (8,4.86) (16,6.65) };

\addplot[color=green!60!black,mark=triangle*,thick] coordinates { (2,1.40) (4,2.44) (8,3.70) (16,4.16) };

\addplot[color=green!60!black,mark=triangle,thick,dashed] coordinates { (2,1.63) (4,2.93) (8,4.87) (16,6.52) };

\nextgroupplot[title={Farm}, ymin=1, ytick={1,2,5,10}, yticklabels={}]
\addplot[color=blue,mark=*,thick] coordinates { (2,1.76) (4,3.40) (8,6.46) (16,9.66) };

\addplot[color=blue,mark=o,thick,dashed] coordinates { (2,1.70) (4,3.41) (8,6.42) (16,8.93) };

\addplot[color=red,mark=square*,thick] coordinates { (2,1.89) (4,3.83) (8,7.34) (16,8.87) };

\addplot[color=red,mark=square,thick,dashed] coordinates { (2,1.63) (4,3.52) (8,6.84) (16,9.01) };

\addplot[color=green!60!black,mark=triangle*,thick] coordinates { (2,1.91) (4,3.61) (8,7.38) (16,11.25) };

\addplot[color=green!60!black,mark=triangle,thick,dashed] coordinates { (2,1.80) (4,3.53) (8,6.85) (16,9.99) };

\nextgroupplot[title={Pipeline}, ymin=1, ytick={1,2,5,10}, yticklabels={}, xlabel={\# threads}]
\addplot[color=blue,mark=*,thick] coordinates { (2,1.61) (4,2.88) (8,4.78) (16,6.28) };

\addplot[color=blue,mark=o,thick,dashed] coordinates { (2,1.86) (4,3.40) (8,5.79) (16,7.33) };

\addplot[color=red,mark=square*,thick] coordinates { (2,1.54) (4,2.71) (8,4.44) (16,5.66) };

\addplot[color=red,mark=square,thick,dashed] coordinates { (2,1.85) (4,3.39) (8,5.94) (16,8.42) };

\addplot[color=green!60!black,mark=triangle*,thick] coordinates { (2,1.39) (4,2.37) (8,3.97) (16,5.61) };

\addplot[color=green!60!black,mark=triangle,thick,dashed] coordinates { (2,1.87) (4,3.41) (8,5.94) (16,8.69) };

\nextgroupplot[title={D\&C}, ymin=1, ytick={1,2,5,10}, yticklabels={}]
\addplot[color=blue,mark=o,thick,dashed] coordinates { (2,1.61) (4,2.50) (8,3.36) (16,3.47) };

\addplot[color=red,mark=square,thick,dashed] coordinates { (2,1.52) (4,2.28) (8,3.17) (16,3.35) };

\addplot[color=green!60!black,mark=triangle,thick,dashed] coordinates { (2,1.51) (4,1.97) (8,2.67) (16,2.73) };

\nextgroupplot[title={MapReduce}, ymin=1, ytick={1,2,5,10}, yticklabels={}]
\addplot[color=blue,mark=*,thick] coordinates { (2,1.70) (4,3.11) (8,5.43) (16,6.08) };

\addplot[color=blue,mark=o,thick,dashed] coordinates { (2,1.86) (4,3.50) (8,6.09) (16,7.77) };

\addplot[color=red,mark=square*,thick] coordinates { (2,1.80) (4,3.28) (8,5.57) (16,8.20) };

\addplot[color=red,mark=square,thick,dashed] coordinates { (2,1.87) (4,3.50) (8,6.56) (16,10.03) };

\addplot[color=green!60!black,mark=triangle*,thick] coordinates { (2,1.48) (4,2.59) (8,4.50) (16,6.35) };

\addplot[color=green!60!black,mark=triangle,thick,dashed] coordinates { (2,1.84) (4,3.53) (8,6.58) (16,9.93) };
\end{groupplot}
\end{tikzpicture}

\vspace{0.5em}
\begin{tikzpicture}[baseline, every node/.style={font=\small, inner sep=2pt}]
    \draw[blue,thick]            (0.0,0.4) -- (0.5,0.4) node[circle, fill=blue, inner sep=1.3pt] {} -- (1.0,0.4);
    \node[anchor=west] at (1.05,0.4) {Medium (100K)};
    \draw[red,thick]             (4.2,0.4) -- (4.7,0.4) node[rectangle, fill=red, minimum size=3pt, inner sep=0pt] {} -- (5.2,0.4);
    \node[anchor=west] at (5.25,0.4) {Large (1M)};
    \draw[green!60!black,thick]  (7.0,0.4) -- (7.5,0.4) node[regular polygon, regular polygon sides=3, fill=green!60!black, minimum size=4pt, inner sep=0pt] {} -- (8.0,0.4);
    \node[anchor=west] at (8.05,0.4) {Extremely Large (5--10M)};
    \draw[thick]                 (0.0,-0.1) -- (1.0,-0.1);
    \node[anchor=west] at (1.05,-0.1) {MP};
    \draw[thick,dashed]          (2.5,-0.1) -- (3.5,-0.1);
    \node[anchor=west] at (3.55,-0.1) {Shared};
  \end{tikzpicture}
\caption{Speedup relative to a plain-JavaScript sequential baseline (main thread, no workers, no SharedArrayBuffer) for all ten patterns across data sizes and thread counts (2--16). The y-axis is log$_{10}$-scaled; the 1x gridline marks the sequential baseline. Lines below 1x indicate that the parallel variant is slower than the sequential loop.}\label{fig:all_speedups}
\end{figure*}

In this section, we evaluate ParaWeb across all ten patterns, covering both CPU variants (MP and Shared) and the GPU variant, in Node.js and in the browser. We first describe the experimental environment, then report CPU scaling for data-parallel patterns and for task-parallel and specialized patterns, followed by browser-based results, GPU acceleration results against the corresponding CPU baseline in each runtime, and a real-time image-convolution case study.

\subsection{Experimental Environment}

Experiments were conducted on an Apple MacBook Pro with the M3~Max chip (16 CPU cores: 12 performance + 4 efficiency; 48~GB unified memory; integrated 40-core Apple GPU), running macOS~26.4, Node.js~v22.19.0, and TypeScript~v5.3.3. GPU benchmarks were executed both under Node.js (using the Dawn-based \texttt{webgpu} npm package) and under headless Chromium (using the native WebGPU implementation over Apple Metal). Each pattern was executed with numeric arrays of the specified sizes, measuring wall-clock execution time. Note that all measurements include two warmup runs to minimize the impact of JIT compilation and cache effects. Results are verified for correctness by comparing outputs across different thread counts to ensure that parallelization does not alter computational results. We additionally collected browser-based results using Web Worker implementations against the same plain-JavaScript sequential baseline with thread counts of 2, 4, 8, and 16, reported at Extremely Large inputs in Table~\ref{tab:webworker_speedup}.

\subsection{Data-Parallel Pattern Results}

Figure~\ref{fig:all_speedups} reports CPU speedup curves for all ten patterns. At 16 threads, Map, Filter, and Scatter achieve comparable speedups of 10.48x, 10.44x, and 11.54x in the Shared variant, and 9.03x, 7.16x, and 5.34x in the MP variant, which indicates that the compute-bound per-element transform amortizes worker-coordination overhead for both variants. Filter's good Shared performance is notable given that variable-length output requires a two-pass count-then-write approach. The savings from avoiding data copying outweigh the synchronization cost. Scatter Shared has the highest 16T speedup of any CPU pattern by writing slice-locally to a shared output buffer and resolving index collisions through a lock-free atomic compare-exchange on a per-output last-writer index, whereas Scatter MP pays for serializing per-worker output chunks back to the main thread.

Reduce and Scan exhibit an arithmetic-intensity threshold: both patterns have a synchronizing combine phase the parallel variants must traverse twice, and below a certain per-element compute cost the coordination overhead dominates. With a per-element trigonometric transform fused into the operator (10 iterations for Reduce, 50 for Scan, matching Map's arithmetic intensity), per-element compute amortizes the coordination cost: Reduce reaches 11.62x (Shared) and 7.77x (MP), and Scan reaches 9.88x and 5.74x respectively at 16 threads. Without that transform, pure associative sum and pure prefix-sum both fall below 1x at every thread count, because the reduction machinery (or for Scan, the second-pass memory traffic) outweighs the trivial per-element work. Therefore, parallel reduction is counterproductive on memory-bandwidth-bound workloads with insufficient per-element compute.

\subsection{Task-Parallel and Specialized Pattern Results}

Farm reaches 11.25x (MP) and 9.99x (Shared) at 16 threads on Node.js, and 11.11x (MP) and 11.85x (Shared) in the browser, on Extremely Large inputs (5M elements; capped below 10M because the per-task workload is heavier than the data-parallel benchmarks). Both variants partition the input into $p$ contiguous ranges via static chunking. The Shared variant additionally pays one \texttt{Atomics.add} per range to support dynamic load balancing. On the Collatz-plus-trigonometric workload, per-task cost variance is mild and static chunking is already balanced, so the dynamic dispatch is unused work, and the two variants converge to within $\sim$10\% of each other in both runtimes. The Shared variant's distinguishing capability translates into measurable benefit only when per-task cost variance is high enough that static chunking creates stragglers, e.g., when heavy tasks cluster in the input array.

On 8M complex points, the Shared D\&C variant reaches 2.73x at 16 threads on Node.js and 2.62x in the browser. The CPU ceiling is intrinsic to FFT: each of the log\textsubscript{2}\,$N \approx 23$ butterfly stages must complete before the next, costing a per-stage coordination round-trip that even Shared cannot avoid. MP cannot scale on FFT at all: per-stage cross-worker data exchange is required, and structured-clone pays for it on every round-trip. As a result, Node MP measures 0.23x at 2 threads (a 4.3x slowdown over sequential) and recovers only to 0.57x at 16 threads, never reaching parity with the sequential baseline. Note that we omit the MP D\&C curves from Figure~\ref{fig:all_speedups} for clarity, since their sub-1x trajectories would compress the readable y-axis range without adding analytical content.

Pipeline, Stencil, and MapReduce all scale predictably with Shared outperforming MP: 8.69x vs.\ 5.61x, 6.52x vs.\ 4.16x, and 9.93x vs.\ 6.35x respectively at 16 threads on Extremely Large inputs. Pipeline benefits from Shared's elimination of inter-stage copying. Stencil's 15-iteration trigonometric refinement per element amortizes the overlap regions duplicated across chunk boundaries. MapReduce mirrors Map's profile because the map phase dominates and the reduce phase adds only a small sequential cost. Within Pipeline, the stream stage operators reach stream-iterate 9.63x, window-reduce 4.12x, and a composed map+filter+window-reduce 5.35x at 16T on 10M-element inputs.

\subsection{Browser-Based Results}

To evaluate ParaWeb in the browser, we re-ran the identical workloads in headless Chromium using the Web Worker variants, served with cross-origin isolation (\texttt{COOP}/\texttt{COEP}) headers to enable \texttt{SharedArrayBuffer}. Table~\ref{tab:webworker_speedup} reports speedups against the same plain-JS sequential baseline used in the Node.js evaluation, at Extremely Large inputs (10M elements; Farm at 5M). Browser speedups follow the Node.js ranking closely: Farm Shared reaches 11.85x at 16 threads, Scatter Shared 11.06x, Farm MP 11.11x, Scan Shared 10.23x, Map Shared 10.02x, MapReduce Shared 9.87x, Filter Shared 9.71x, Reduce Shared 9.43x, Pipeline Shared 8.01x, and Stencil Shared 6.10x. D\&C reaches 2.62x (Shared) and 2.11x (MP) on the FFT workload, limited by the per-stage synchronization shared by all variants.

\begin{table}[!htbp]
\caption{Web Worker speedup (relative to a plain-JavaScript sequential baseline: main thread, no workers, no \texttt{SharedArrayBuffer}) for Extremely Large inputs (10M elements; Farm at 5M to match its heavier per-task workload, D\&C at 8M to keep $N$ a power of two for the radix-2 FFT), measured in headless Chromium. Bold values mark the per-pattern peak; italic values below 1.0x indicate the parallel variant is slower than the sequential loop.}\label{tab:webworker_speedup}
\newcommand{\thead}[1]{\rotatebox[origin=l]{90}{#1}}
\begin{tabular*}{\textwidth}{@{\extracolsep{\fill}}llrrrrrrrrrr@{}}
\toprule
T & Variant & \thead{Map} & \thead{Filter} & \thead{Reduce} & \thead{Scan} & \thead{Scatter} & \thead{Stencil} & \thead{Farm} & \thead{Pipeline} & \thead{D\&C} & \thead{MapReduce} \\
\midrule
\multirow{2}{*}{2}  & MP     & 1.71 & 1.53 & 1.12 & 1.35 & 1.44 & 1.14 & 1.85 & 1.39 & 1.11 & 1.60 \\
                    & Shared & 1.89 & 1.87 & 1.93 & 1.87 & 1.91 & 1.38 & 1.85 & 1.81 & 1.53 & 1.86 \\
\midrule
\multirow{2}{*}{4}  & MP     & 3.05 & 2.64 & 1.84 & 2.36 & 2.46 & 1.88 & 3.62 & 2.37 & 1.97 & 2.89 \\
                    & Shared & 3.58 & 3.53 & 3.74 & 3.52 & 3.50 & 2.52 & 3.66 & 3.29 & 2.15 & 3.51 \\
\midrule
\multirow{2}{*}{8}  & MP     & 5.30 & 4.25 & 2.79 & 3.84 & 3.63 & 2.92 & 6.91 & 3.72 & 2.21 & 4.79 \\
                    & Shared & 6.66 & 6.33 & 6.88 & 6.72 & 6.86 & 4.35 & 6.98 & 5.64 & 2.57 & 6.49 \\
\midrule
\multirow{2}{*}{16} & MP     & 7.82 & 5.92 & 3.43 & 5.16 & 4.39 & 3.69 & 11.11 & 5.29 & 2.11 & 7.18 \\
                    & Shared & \textbf{10.02} & \textbf{9.71} & \textbf{9.43} & \textbf{10.23} & \textbf{11.06} & \textbf{6.10} & \textbf{11.85} & \textbf{8.01} & \textbf{2.62} & \textbf{9.87} \\
\botrule
\end{tabular*}
\end{table}

\subsection{GPU Acceleration Results}

We evaluate the GPU variants of all ten patterns in both runtimes, using the same data sizes and equivalent workloads as the CPU benchmarks, and compare GPU execution against the corresponding CPU variant within the same runtime: Worker-Thread CPU in Node.js and Web Worker CPU in the browser. Since the GPU handles parallelism internally through its hardware scheduler, there is no thread count parameter. Each benchmark measures the total execution time, including data upload, kernel execution, and result readback. Table~\ref{tab:gpu_comparison} reports the Node.js and browser GPU results side by side.

\begin{table}[!htbp]
\caption{GPU vs.\ CPU performance on Extremely Large inputs (10M elements; Farm at 5M and D\&C at 8M to match CPU caps) for both runtimes. Speedups are reported against the plain-JS sequential baseline (``vs seq'') and against the best parallel CPU variant at 16 threads (``vs 16T''). Bold marks the peak per column; italic values below 1.0x indicate CPU beats GPU. All rows compare equivalent workloads: Reduce, Scan, Scatter, Farm, and D\&C GPU kernels apply the same per-element transform or algorithm as their CPU counterparts.}\label{tab:gpu_comparison}
\begin{tabular*}{\textwidth}{@{\extracolsep{\fill}}lrrrrrr@{}}
\toprule
& \multicolumn{3}{c}{Node.js} & \multicolumn{3}{c}{Browser} \\
\cmidrule(lr){2-4}\cmidrule(lr){5-7}
Pattern & GPU (ms) & vs seq & vs 16T & GPU (ms) & vs seq & vs 16T \\
\midrule
Map                 & 163 & 63.5x           &  6.1x           & 176 & 54.7x           &  5.5x \\
Filter              & 950 &  8.0x           & \textit{0.8x}   & 969 &  6.5x           & \textit{0.7x} \\
Reduce              &  22 & 94.2x           &  8.1x           &  30 & 46.4x           &  4.9x \\
Scan                & 222 & 67.0x           &  6.8x           & 191 & 65.2x           &  6.4x \\
Scatter             &  85 & 239.6x          & 20.8x           & 114 & 113.5x          & 10.3x \\
Stencil             & 160 & 17.7x           &  2.7x           & 164 & 16.4x           &  2.7x \\
Farm                & 105 & \textbf{260.5x} & \textbf{22.9x}  & 108 & \textbf{222.6x} & \textbf{21.9x} \\
Pipeline            & 324 & 25.7x           &  3.0x           & 324 & 20.6x           &  2.6x \\
D\&C                & 479 &  3.6x           &  1.3x           & 423 &  3.0x           &  1.1x \\
MapReduce           & 193 & 53.7x           &  5.4x           & 191 & 49.8x           &  5.1x \\
\botrule
\end{tabular*}
\end{table}

The GPU outperforms 16-thread CPU on data- and task-parallel patterns with high per-element arithmetic intensity, with the largest Node.js speedups on Farm (260.5x), Scatter (239.6x), Reduce (94.2x), Scan (67.0x), Map (63.5x), MapReduce (53.7x), Pipeline (25.7x), and Stencil (17.7x), with speedups of 3--23x over the best 16-thread CPU configuration for most patterns. The browser column mirrors this ranking with lower ratios because both CPU and GPU share the Apple Metal driver through Chromium's scheduler. Filter loses to CPU-16T in both runtimes because V8 JIT-optimizes its CPU path well~\cite{martinsen2017v8} and the three-phase mark/scan/compact GPU kernel is dominated by the sequential single-workgroup prefix-sum in compaction. D\&C reaches a GPU speedup of 3.6x in Node.js and 3.0x in the browser, which is bounded by FFT's inherent cross-stage synchronization: the log\textsubscript{2}\,$N$ stages must execute in order even though the pairwise operations within a stage are independent.

\subsection{Case Study: Real-Time Image Convolution}\label{subsec:casestudy}

To show that ParaWeb is useful for a concrete web scenario, we present a case study of 2D image convolution, which is a core building block in privacy-preserving browser-based applications such as photo editors, medical imaging viewers, and real-time video filters. In such applications, user data remains on the client device and computation must complete within the typical interactive latency budget of approximately 100~ms per operation for slider-based filtering to feel responsive. We evaluate five common convolution filters: Gaussian blur, box blur, unsharp-mask sharpen, directional emboss, and Difference-of-Gaussians edge detection~\cite{marr1980edge}.

\begin{figure*}[!htbp]
\centering
\begin{tikzpicture}
\begin{groupplot}[
  group style={
    group size=5 by 1,
    horizontal sep=4pt,
    xticklabels at=edge bottom,
    yticklabels at=edge left,
  },
  width=3.9cm,
  height=5.2cm,
  ymode=log,
  log basis y=10,
  ytick={1,2,5,10,20,50,100,200,500},
  yticklabels={1x,2x,5x,10x,20x,50x,100x,200x,500x},
  ymin=2, ymax=500,
  xmin=0, xmax=21,
  xtick={1,2,4,8,12,16,20},
  xticklabels={1,2,4,8,12,16,20},
  grid=major,
  title style={at={(axis description cs:0.04, 0.95)}, anchor=north west, font=\scriptsize, fill=white, inner sep=1.5pt},
  tick label style={font=\footnotesize},
  label style={font=\footnotesize},
  every axis x label/.style={at={(axis description cs:0.5, -0.22)}, anchor=north},
]
\nextgroupplot[title={Gaussian}]
\addplot[color=orange!70!yellow,mark=o,thick] coordinates { (1,3.04) (2,3.29) (4,3.41) (8,3.56) (12,3.69) (16,3.47) (20,3.60) };
\addplot[color=orange!90!black,mark=square,thick] coordinates { (1,4.69) (2,5.40) (4,5.89) (8,6.54) (12,6.89) (16,6.62) (20,6.92) };
\addplot[color=red!80!black,mark=triangle,thick] coordinates { (1,5.07) (2,5.99) (4,6.91) (8,8.24) (12,9.61) (16,9.12) (20,8.69) };
\addplot[color=blue!60!black,mark=*,thick] coordinates { (1,4.99) (2,5.83) (4,11.58) (8,20.13) (12,29.40) (16,38.56) (20,49.58) };

\nextgroupplot[title={Box}]
\addplot[color=orange!70!yellow,mark=o,thick] coordinates { (1,3.27) (2,3.34) (4,3.43) (8,3.50) (12,3.51) (16,3.48) (20,3.43) };
\addplot[color=orange!90!black,mark=square,thick] coordinates { (1,4.94) (2,5.63) (4,5.99) (8,6.44) (12,5.26) (16,6.37) (20,6.64) };
\addplot[color=red!80!black,mark=triangle,thick] coordinates { (1,5.23) (2,6.15) (4,6.82) (8,7.79) (12,8.94) (16,9.34) (20,9.54) };
\addplot[color=blue!60!black,mark=*,thick] coordinates { (1,5.77) (2,8.09) (4,12.74) (8,21.56) (12,28.54) (16,38.73) (20,47.66) };

\nextgroupplot[title={Sharpen}, xlabel={kernel radius}]
\addplot[color=orange!70!yellow,mark=o,thick] coordinates { (1,3.42) (2,3.48) (4,3.50) (8,3.55) (12,3.49) (16,3.61) (20,3.57) };
\addplot[color=orange!90!black,mark=square,thick] coordinates { (1,5.35) (2,5.70) (4,6.15) (8,6.50) (12,6.59) (16,6.88) (20,6.89) };
\addplot[color=red!80!black,mark=triangle,thick] coordinates { (1,5.61) (2,6.09) (4,7.02) (8,7.30) (12,8.84) (16,9.81) (20,8.07) };
\addplot[color=blue!60!black,mark=*,thick] coordinates { (1,6.50) (2,8.55) (4,13.57) (8,22.00) (12,29.95) (16,42.35) (20,49.35) };

\nextgroupplot[title={Emboss}]
\addplot[color=orange!70!yellow,mark=o,thick] coordinates { (1,2.94) (2,3.35) (4,3.66) (8,3.78) (12,3.40) (16,3.78) (20,3.68) };
\addplot[color=orange!90!black,mark=square,thick] coordinates { (1,4.87) (2,5.80) (4,6.92) (8,7.22) (12,6.37) (16,6.73) (20,6.25) };
\addplot[color=red!80!black,mark=triangle,thick] coordinates { (1,5.43) (2,7.83) (4,10.57) (8,11.31) (12,10.16) (16,11.21) (20,9.73) };
\addplot[color=blue!60!black,mark=*,thick] coordinates { (1,6.91) (2,15.43) (4,40.09) (8,123.91) (12,226.80) (16,314.17) (20,414.42) };

\nextgroupplot[title={Edge (DoG)}]
\addplot[color=orange!70!yellow,mark=o,thick] coordinates { (1,3.81) (2,3.79) (4,3.70) (8,2.90) (12,3.57) (16,3.63) (20,3.65) };
\addplot[color=orange!90!black,mark=square,thick] coordinates { (1,6.62) (2,6.88) (4,6.94) (8,7.04) (12,7.07) (16,7.07) (20,7.11) };
\addplot[color=red!80!black,mark=triangle,thick] coordinates { (1,7.45) (2,7.44) (4,8.34) (8,9.41) (12,9.91) (16,9.74) (20,9.89) };
\addplot[color=blue!60!black,mark=*,thick] coordinates { (1,12.00) (2,19.67) (4,32.21) (8,56.70) (12,70.65) (16,79.49) (20,85.06) };
\end{groupplot}
\end{tikzpicture}

\vspace{0.3em}
\begin{tikzpicture}[baseline, every node/.style={font=\small, inner sep=2pt}]
    \draw[orange!70!yellow,thick] (0.0,0) -- (0.5,0) node[circle, draw=orange!70!yellow, fill=orange!70!yellow, inner sep=1.5pt] {} -- (1.0,0);
    \node[anchor=west] at (1.05,0) {Shared 4T};
    \draw[orange!90!black,thick] (2.7,0) -- (3.2,0) node[rectangle, draw=orange!90!black, fill=orange!90!black, minimum size=3pt, inner sep=0pt] {} -- (3.7,0);
    \node[anchor=west] at (3.75,0) {Shared 8T};
    \draw[red!80!black,thick] (5.4,0) -- (5.9,0) node[regular polygon, regular polygon sides=3, draw=red!80!black, fill=red!80!black, minimum size=4pt, inner sep=0pt] {} -- (6.4,0);
    \node[anchor=west] at (6.45,0) {Shared 16T};
    \draw[blue!60!black,thick] (8.3,0) -- (8.8,0) node[circle, draw=blue!60!black, fill=blue!60!black, inner sep=1.3pt] {} -- (9.3,0);
    \node[anchor=west] at (9.35,0) {GPU};
  \end{tikzpicture}
\caption{Image-convolution case study (3840x2160): speedup relative to the sequential CPU baseline as the kernel radius grows, for each of the five filters. Y-axis is log$_{10}$-scaled.}\label{fig:casestudy_scaling}
\end{figure*}
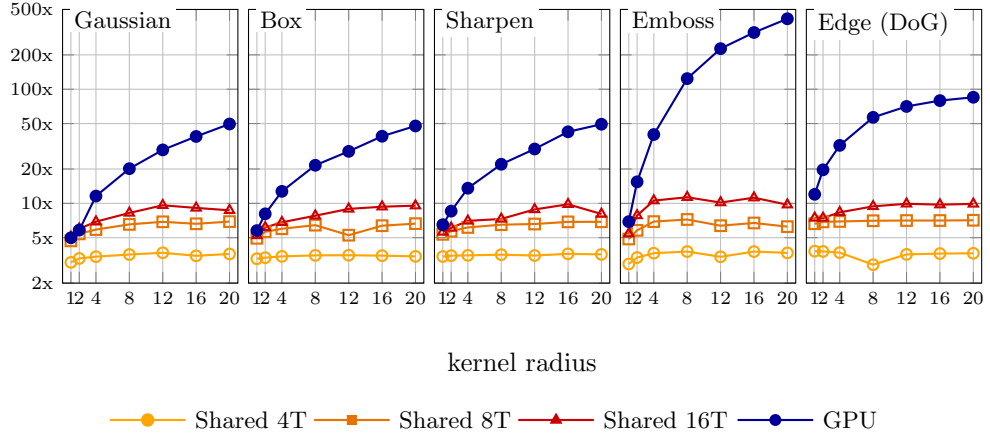

Each filter maps onto one or two of ParaWeb's patterns. Gaussian blur and box blur decompose into two 1D Stencil passes (horizontal followed by vertical), exploiting kernel separability to reduce per-pixel work from $O(r^2)$ to $O(r)$. Emboss uses a single 2D Stencil with a full $(2r+1) \times (2r+1)$ non-separable kernel. Unsharp-mask sharpen composes a Stencil (blur) with a Map that combines source and blurred pixels element-wise ($\textit{out} = \textit{src} + (\textit{src} - \textit{blur})$). DoG edge detection composes two Stencils at different radii with a Map that subtracts and biases. Sharpen and DoG are therefore naturally expressed as short Pipelines. Our case-study implementation uses a specialized worker on top of ParaWeb's worker pool and \texttt{SharedArrayBuffer} utilities that follows this same structure, and the GPU path is realized as one or two fused WGSL compute shaders (16$\times$16 workgroups) per filter so intermediate buffers stay on the device. Both variants are measured in the steady state of an interactive session: the CPU path reuses its worker pool and shared input/output buffers across calls, and the GPU path caches its compiled pipeline, device buffers, and bind group. Per-call work is the image upload plus compute. Figure~\ref{fig:casestudy_scaling} plots speedup relative to the sequential CPU baseline as the kernel radius grows from 1 to 20 on a 4K image, on a log y-axis.

The separable filters (Gaussian, box) and the composite unsharp-mask sharpen share the same scaling profile: GPU speedup climbs from around 5x at radius~1 to 40--50x at radius~20, while CPU-Shared 16T plateaus around 8--10x because memory bandwidth rather than compute dominates. The non-separable emboss filter, whose cost scales as $(2r+1)^2$ per pixel, places significantly higher demand on the parallel infrastructure: the GPU reaches a 414x speedup at radius~20 (75~ms versus 31~s sequential) because its 5{,}120 parallel lanes (40 Apple GPU cores $\times$ 128 ALUs) absorb the quadratic work that saturates even the 16-thread CPU configuration at ~10x. This exceeds the GPU speedups in Table~\ref{tab:gpu_comparison} for the natively transcendental-heavy patterns (e.g., 260x on Farm, 240x on Scatter, 64x on Map) because emboss at radius~20 is 1681 fused multiply-adds per pixel with no transcendentals, matching the GPU's FMA lanes almost perfectly, whereas the pattern microbenchmarks are intentionally transcendental-heavy and thus bound by the GPU's special-function units rather than its FMA throughput. Edge detection via DoG, which is two separable blurs plus a per-pixel combine, falls between these regimes and reaches 85x on the GPU. Note that the same sweep on $1024^2$ and $2048^2$ images yields the same ranking and almost-identical GPU-to-CPU ratios, with the emboss GPU peak at radius~20 staying in the 380--414x band across all three resolutions. This confirms that the scaling shape is a property of the algorithms rather than of a specific image size. This case study demonstrates that for compute-bound browser workloads such as client-side image and video processing, ParaWeb shifts applications from batch-style processing with visible latency to genuinely interactive behavior without sending user data to a server.


\subsection{Discussion}

Across configurations, we observe several trends. First, on Node.js the Shared variant outperforms MP on most compute-bound patterns, reaching 9--12x at 16 threads (Figure~\ref{fig:all_speedups}). Farm is the one exception, where Shared (9.99x) trails MP (11.25x) because static chunking already balances the workload and the atomic counter adds unused coordination cost. Second, the variant ordering is not stable across runtimes: in the browser the ranking flips for Farm, with Shared (11.85x) overtaking MP (11.11x) (see Table~\ref{tab:webworker_speedup}). Third, Reduce and Scan are memory-bandwidth bound with a pure-associative workload, so parallel scaling is limited or negative when no per-element compute is added. Both patterns scale past 9x once a per-element transform is fused into the operator.

These results indicate that variant defaults are workload-dependent rather than fixed per pattern. Reduce and Scan illustrate this in particular: both fall below 1x without a per-element transform but scale past 9x once one is fused into the operator, so the appropriate variant depends on arithmetic intensity rather than on the pattern alone. D\&C illustrates a complementary case: it plateaus at $\sim$2.7x because Cooley-Tukey FFT's cross-stage synchronization is intrinsic to the algorithm rather than a property of any one variant, and the appropriate strategy for FFT-shaped workloads is therefore the GPU rather than a different CPU variant. ParaWeb's static default selection (Shared for most patterns on CPU, with GPU recommended for compute-bound patterns with high arithmetic intensity such as Farm, Scatter, Map, MapReduce, Pipeline, and Stencil) is therefore a useful starting point, while a workload-aware selector is a longer-term direction.

The three variants reflect complementary trade-offs. MP is simpler but incurs copying. Shared eliminates copying but adds coordination complexity and restricts data to typed arrays. GPU offers the highest throughput for high-arithmetic-intensity patterns but requires WGSL and \texttt{f32} precision. Current limitations include cross-origin isolation for browser Shared variants, \texttt{f32} GPU precision, a sequential single-workgroup prefix-sum inside GPU Filter's compaction step (the standalone Scan pattern already uses a parallel Blelloch kernel~\cite{blelloch1990scan}), the requirement that user-supplied functions be self-contained because \texttt{new Function()} in the worker does not preserve the caller's lexical scope, and static variant selection.

\section{Conclusion and Future Work}\label{sec:conclusion}

In this paper, we presented ParaWeb, a TypeScript library that implements ten parallel programming patterns for Node.js, browsers, and WebGPU, with three implementation variants per pattern. CPU variants reach speedups of up to 11.6x in Node.js at 16 threads on compute-bound patterns. GPU variants reach 260x (Farm) on Node.js for patterns with high arithmetic intensity. Our image convolution case study reaches 414x GPU speedup on the non-separable emboss filter at radius~20. Future work may focus on dynamic runtime variant selection, fused GPU kernels for composed patterns, and heterogeneous workload distribution that splits a single pattern invocation across CPU workers and the GPU concurrently rather than picking one. A complementary direction is a cloud-continuum-aware scheduler that decides per call whether to execute on client resources, offload to a server runtime, or split work between the two based on input size, network conditions, and data residency constraints.

\begingroup
\footnotesize
\setlength{\bibsep}{0pt}
\setlength{\itemsep}{0pt}
\setlength{\parskip}{0pt}

\endgroup

\end{document}